\documentclass[aps,twocolumn,showpacs]{revtex4}
\usepackage{dcolumn}
\usepackage{float}
\usepackage{graphicx}
\usepackage{amsmath}
\usepackage{booktabs}
\usepackage{array}
\usepackage{multirow}
\usepackage{makecell}
\usepackage{verbatim}
\usepackage{amssymb}
\usepackage{psfrag}
\usepackage{color}
\usepackage{bm}
\usepackage{wrapfig}
\usepackage{subfigure}
\usepackage{makeidx}
\usepackage{bm}
\usepackage{epsf}
\usepackage{multirow}
\usepackage{mathrsfs}
\usepackage{braket}
\usepackage[colorlinks,linkcolor=blue,urlcolor=blue,citecolor=blue]{hyperref}

\begin{document}
\title{Anomalous Trajectory Drift and Geometric Phases of Cyclic Spinor Solitons Induced by Virtual Magnetic Monopoles}
\author{Ruo-Yun Wu$^{1}$}
\thanks{These authors contribute equally.}
\author{Ning Mao$^{1}$}
\thanks{These authors contribute equally.}
\author{Xiao-Lin Li$^{1}$}
\author{Jie Liu$^{2}$}
\author{Li-Chen Zhao$^{1,3,4}$}\email{zhaolichen3@nwu.edu.cn}

\address{$^{1}$School of Physics, Northwest University, Xi'an, 710127, China}
\address{$^{2}$Graduate School, China Academy of Engineering Physics, Beijing 100193, China}
\address{$^{3}$NSFC-SPTP Peng Huanwu Center for Fundamental Theory, Xi'an 710127, China}
\address{$^{4}$Shaanxi Key Laboratory for Theoretical Physics Frontiers, Xi'an 710127, China}
\date{\today}
\begin{abstract}
We investigate the dynamics of a two-component Bose-Einstein condensate  with spin-orbit coupling numerically and analytically. Under the drive of a weak segmented rotational external field, we observe that the system exhibits cyclic soliton motion; however, in contrast to the predictions of quasi-particle theory, the trajectory of the soliton center shows a distinct drift. The underlying mechanism of this anomalous drift is revealed: the moving soliton experiences a Lorentz force induced by a virtual magnetic  monopole field in momentum space. We further calculate the phase evolution of the soliton during this cyclic motion and find that its geometric component comprises both an adiabatic Berry phase and a nonadiabatic Aharonov-Anandan phase. Notably, the Berry phase can be expressed in terms of the magnetic flux of the aforementioned virtual monopole field. Our findings hold implications for geometric phase theory and experiments on two-component Bose-Einstein condensates, and may establish a novel link between quantum geometry and soliton dynamics.

\end{abstract}
\maketitle

\emph{Introduction.}
Solitons are localized wave packets that exhibit inherent stability, arising from a delicate balance between nonlinear self-interaction and diffusion induced by intrinsic dispersion \cite{soliton,Guo}.
The emergence of solitons signals ordered self-organization in complicated dynamical systems, making their study pivotal to integrability theory \cite{darboux1,darboux2,JYang}.
Solitons also hold substantial applications in many different fields \cite{Redondo}, such as  information storage \cite{Prilepsky,Nolan}, transport \cite{Mostaan,Ye1,Jurgensen,Jurgensen2}, quantum sensors \cite{McDonald,Helm,Grimshaw}, and quantum computations \cite{Mukherjee,Amico}.

Bose-Einstein condensates (BECs), a recently realized state of matter, represent one of the most suitable platforms for investigating soliton dynamics. This is attributed to their high flexibility in engineering and manipulating key physical parameters, such as nonlinear interaction strengths and external potentials  \cite{Stringari,34xde,Kartashov}. Varied kinds of solitons predicted by integrability theory  \cite{46DB,cnse,56Vary1,55spin},  along with their extended forms (e.g., soliton-vortex complexes), have been experimentally observed in BECs in recent years \cite{3DB,5Magnetic,35DDE1,36DDE2,VortexE,VortexE1,VortexE2}.
Owing to their quasi-classical particle-like properties, the dynamical evolution of solitons can, in most cases, be well described by quasi-particle theory (QPT). This theory is founded on the principles of energy conservation and the momentum theorem, with the approximation of neglecting the spatial extension of the soliton wave packet \cite{Busch,Lewenstein,Qu,Scott,Brand}.

Coupled multi-component Bose-Einstein condensates (BECs) have attracted considerable interest in recent research \cite{39SP,Zhang2016,Manchon2015}.  One of their important applications is to simulate spin-orbit coupled (SOC) effects in varied physical systems, which have exhibited many exotic phenomena \cite{Li2015,Lin2011,Galitski2013,Pan2014,Huang2016}. BECs with spin-orbit coupling (SOC-BECs) exhibit novel ground states, including supersolids and stripe phases \cite{HZhai2,HZhai,Ssolid}, and support a range of nonlinear excitations such as stripe solitons \cite{Kevrekidis2013,Mithun2024,Xu2013} and high-dimensional solitons \cite{YCZhang,Boris}. When subjected to a constant driving force, multi-component solitons undergo Josephson-like oscillations, which originate from a negative-to-positive mass transition  \cite{Zhao2020,Yu,MengPRA2022,Bresolin,Mengzhao}. This remarkable phenomenon has been recently observed experimentally \cite{Rabec2025}.

In this work,  we investigate the dynamics of a two-component SOC-BEC driven by an external field, aiming to address how the evolutionary dynamics of BEC solitons are influenced by the interplay between SOC effects and external field driving.  When the external field is of a segmented rotational form, we observe that the system supports the intriguing cyclic motion of a localized BEC wave packet. However, in contrast to QPT predictions, the orbit of the wave packet center exhibits a distinct drift. To analyze this phenomenon, we introduce a trial wave function constructed as the product of a spatially dependent soliton profile, a propagating plane wave, and spatially independent spinor components. Leveraging the Lagrangian variation principle, we identify that a Lorentz force generated by a virtual magnetic monopole (VMM) field in momentum space, accounts for the observed anomalous trajectory drift. Furthermore, we find that the VMM simultaneously modulates the phase evolution of the soliton during its cyclic motion.

\begin{figure}
    \centering
    \includegraphics[width=1\linewidth]{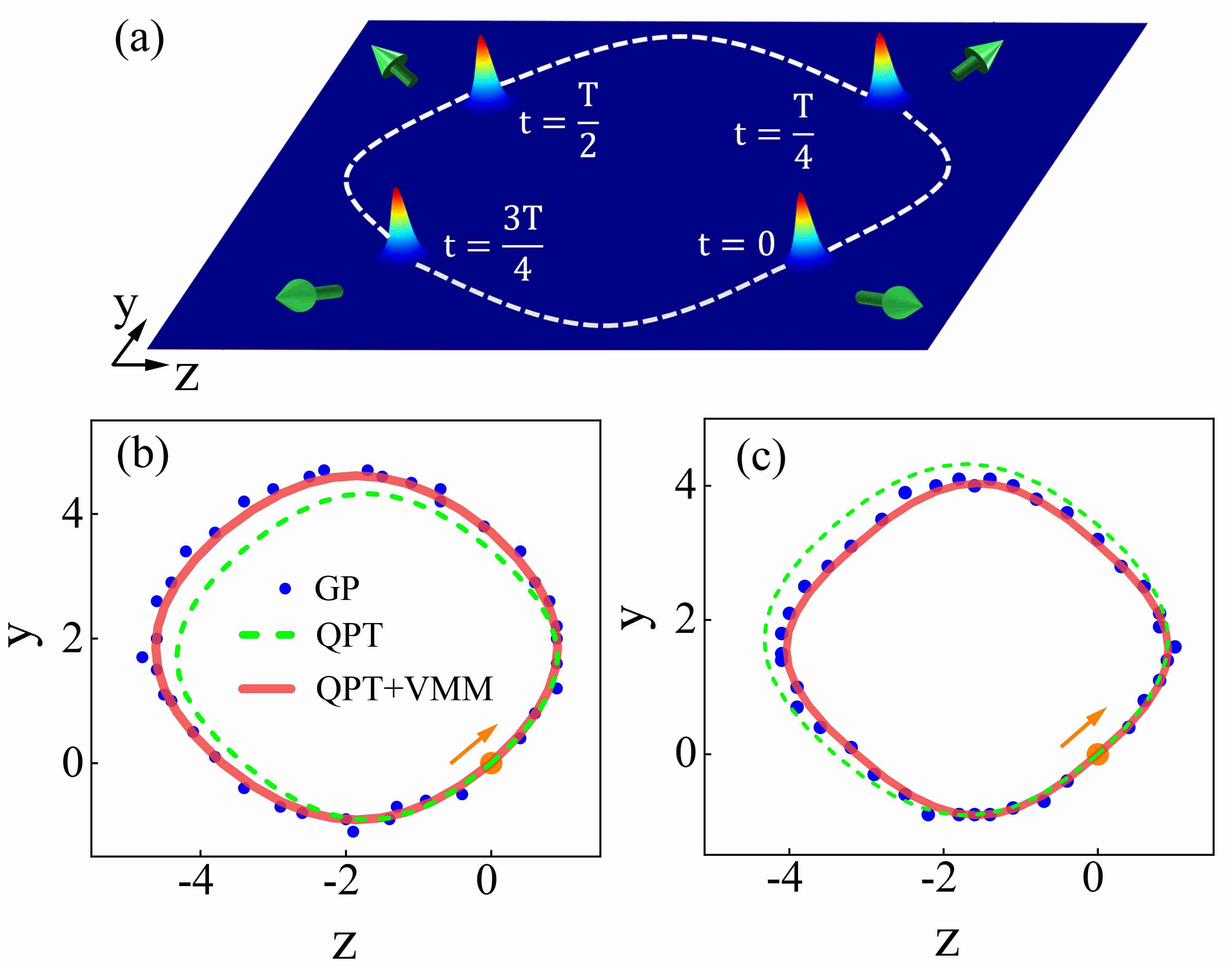}
     \caption{ (a) Schematic plotting  of the wavepacket  motion as well as the rotation of spinor components (green arrows). (b-c) The trajectories of soliton center for the cases with $\Omega=1$ and  $-1$ respectively. For each case,  the trajectory from GP is marked by blue dots, the QPT is denoted by green dashed line, and red solid line is the one predicted by QPT+VMM. Details refer to maintext. }
    \label{fig1}
\end{figure}

\emph{Anomalous soliton trajectory drift.}
We consider a SOC two-component BEC driven by an external field, which is governed by the following dimensionless Gross-Pitaevskii (GP) equations \cite{Kevrekidis2013,Mithun2024,Xu2013}
$
i \frac{\partial  }{\partial t} \begin{pmatrix}
\psi_{\uparrow} \\
\psi_{\downarrow}
\end{pmatrix} = [-\frac{1}{2}\nabla^2  + V(\mathbf{r},t)
+ g\left(|\psi_{\downarrow}|^2 + |\psi_{\uparrow}|^2\right) ] \begin{pmatrix}
\psi_{\uparrow} \\
\psi_{\downarrow}
\end{pmatrix} +(- i \frac{\partial }{\partial z} \sigma_z - i \frac{\partial  }{\partial y} \sigma_y + \Omega \sigma_x)  \begin{pmatrix}
\psi_{\uparrow} \\
\psi_{\downarrow}
\end{pmatrix}$,
where $\psi_{\uparrow,\downarrow}(r,t)$ represent the wavefunctions of the two components, respectively. $\nabla^2$ is the Laplacian accounting for kinetic energy, and
the linear spatial derivative terms, i.e., $p_z \sigma_z+p_y \sigma_y$ describe a SOC.  The SOC can be realized by two Raman laser beams with a frequency difference matching the transition frequency of atomic internal states that irradiate atoms at a specific angle to induce the transitions between different hyperfine states (denoted by $\psi_{\uparrow}$ and $\psi_{\downarrow}$) of the atoms, accompanied by momentum transfer \cite{Huang2016,Lin2011,Galitski2013}. $\Omega \sigma_x$ denotes the Rabi coupling between the two components.  $V(\mathbf{r},t)$ is a time-dependent external potential. The parameter $g$ describe the nonlinear self-interaction strength in a mean-field approximation.

The above GP equations are nonlinear and non-integrable. We can exploit numerical algorithm of the Fourier pseudo-spectral method with Strang splitting \cite{JYang,BaoKRM2013} to trace the evolution of the BEC. The initial state of the BEC takes a localized wavepacket form expressed explicitly by  $
\psi(\mathbf{r} )=\begin{pmatrix}
\psi_{\uparrow} \\
\psi_{\downarrow}
\end{pmatrix} = a
\operatorname{sech}\left[\mathrm{w} |\mathbf{r}-\mathbf{r}_c |\right] e^{i\mathbf{k}  \cdot (\mathbf{r} - \mathbf{r}_c)} \begin{pmatrix}
C_1 \\
C_2
\end{pmatrix} $.
 In our simulations, the amplitude of soliton $a=0.116$ , the width $\mathrm{w}=0.1 $, the nonlinear parameter $g=-1$, the initial momentum vector $\mathbf{k}  = (k_{z0}=1 , k_{y0}=1 )$ and initial position $\mathbf{r}_c=(z_{c0}=0,y_{c0}=0)$. $C_1=\frac{-1+i} {\sqrt{6+2 \sqrt{3}}}$ and $C_2=\frac{1+\sqrt{3}}{\sqrt{6+2 \sqrt{3}}}$ for the case with $\Omega=1$ determine the initial soliton amplitudes of the two components.
An external potential  $V(\mathbf{r},t)$ is rotational, consisting of piecewise linear segments with period $T=40$:
$V(\mathbf{r},t) = F z $ for    $ t \in (0,\frac{ T}{4}]$,  $F y   $ for $ t \in (\frac{ T}{4}, \frac{ T}{2}]$,
$-F z $ for   $t \in (\frac{ T}{2}, \frac{ 3T}{4}]$,  and $-F y $  for  $t \in ( \frac{ 3T}{4}, T]$.

The time evolution of the initial wavepacket with $\Omega=1$ is presented in Fig. \ref{fig1} (a), where the four subfigures display the density distributions of the wavepacket at times $t=0, T/4, T/2$ and $3T/4$ respectively. It can be observed that the wavepacket remains spatially localized throughout the evolution. More notably, under the rotational external field, the wavepacket returns perfectly to its initial position after one period. A rotational external force typically does not ensure cyclic motion of the driven particle. However, in our case, by appropriately selecting the initial settings of the soliton and the force duration based on the strength of the applied external force, i.e., $8 k_{z0}/T = 8k_{y0}/T = F=0.2$, we achieve cancellation between the positive and negative displacements of the soliton. During the cyclic motion of the soliton's mass density, its spin vector also undergoes periodic temporal variations, as schematically indicated by the green arrows in Fig. \ref{fig1} (a).

Since solitons usually behaves like a particle, the QPT can be conveniently used to analyze its motion \cite{Busch,Lewenstein,Qu,Scott,Brand,Zhao2020,Yu,MengPRA2022,Bresolin,Mengzhao}. With using the trial wave function of soliton $\psi(\mathbf{r})$,  the soliton excitation energy can be calculated,
$E_s(\mathbf{k}) = \int \bigg[ \frac{1}{2} |\nabla \psi_\uparrow|^2 + \frac{1}{2} |\nabla \psi_\downarrow|^2   - \psi_\uparrow^*(i\partial_z\psi_\uparrow + \partial_y\psi_\downarrow)+ \psi_\downarrow^*(i\partial_z\psi_\downarrow + \partial_y\psi_\uparrow) +\psi_\uparrow^*\Omega\psi_\downarrow+\psi_\downarrow^*\Omega\psi_\uparrow + \frac{g}{2} (|\psi_\uparrow|^4 + |\psi_\downarrow|^4)  + g|\psi_\uparrow|^2 |\psi_\downarrow|^2 \bigg] d\mathbf{r} =  N_b ( \frac{k_z^2+k_y^2}{2}-\sqrt{k_z^2+ k_y^2+\Omega^2})+(\frac{\ln 2}{3}+\frac{1}{6} ) \pi a^2 +  (\frac{2\ln 2}{3}-\frac{1}{6}) \frac{\pi g a^4}{\mathrm{w}^2} $.
In the above deductions, the adiabatic approximation is employed: we consider a scenario where the external force is small compared to the spin excitation energy, i.e., $F \ll \sqrt{k_z^2+k_y^2+\Omega^2} $. Under this condition, the BEC spinor remains in an eigenstate of the SOC Hamiltonian excluding the nonlinear terms.  We can thus derive the spinor parameters as $C_1 =\frac{ik_y-\Omega} {\sqrt{2(k_z^2+k_y^2+\Omega^2)+2 k_z \sqrt{k_z^2+k_y^2+\Omega^2} }}$ and $C_2= \frac{k_z+ \sqrt{k_z^2+k_y^2+\Omega^2}} {\sqrt{2(k_z^2+k_y^2+\Omega^2)+2 k_z \sqrt{k_z^2+k_y^2+\Omega^2}}}$ (see details in \cite{SM}). Here we assume the spinor parameters are independent of spatial variables, since we consider the weak nonlinear interactions.

The potential energy generated by the weak segmented rotational external field is given by:
$ E_p(\mathbf{r}_c)= \int V(\mathbf{r},t) ({|\psi_\uparrow |}^2+{|\psi_\downarrow |}^2) d\mathbf{r}
      = N_b V(\mathbf{r}_c,t)$. In the above expressions, $N_b=\int {|\psi_\uparrow |}^2+{|\psi_\downarrow |}^2 d\mathbf{r}=\frac{2\pi \ln 2 a^2}{\mathrm{w}^2} $. Based on the momentum theorem $\mathbf{F}=\dot{\mathbf{p}}$, where
the soliton momentum is $\mathbf{p} = \frac{i}{2}\int (\psi\nabla\psi^* - \psi^*\nabla\psi)d\mathbf{r}=N_b \mathbf{k}$ and
the external force is $\mathbf{F}=-\nabla  E_p(\mathbf{r}_c)=-N_b \nabla V(\mathbf{r}_c,t)$, we obtain $\dot {\mathbf{k}}=-\nabla V(\mathbf{r}_c,t)$.
For each of the four steps in one period, energy conservation holds: $E_s(\mathbf{k})+E_p(\mathbf{r}_c)= const $, which yields $\dot{\mathbf{r}}_c =  \frac{1}{N_b} \nabla_{\mathbf{k}} E_s(\mathbf{k})$. The calculated soliton trajectories for \(\Omega=1,-1\) are shown by the green lines in Fig. \ref{fig1} (b) and (c), demonstrating that the QPT can successfully predict cyclic soliton motion under segmented rotational external fields. However, the observed trajectories from GP equations have  apparent drifts compared to the predictions of QPT  and only coincide with numerical results in the vicinity of the initial position.  As shown below, only with considering the contribution of VMM field, these anomalous trajectory drifts can be explained properly.

\emph{Theoretical analysis of the anomalous drift.}
The SOC two-component BEC governed by the GP equations is derived from the time-dependent variational principle applied to the following Lagrangian: $L = \int \psi^{\dag}  i\partial_t   \psi   - [ \frac{1}{2} |\nabla \psi_\uparrow|^2 + \frac{1}{2} |\nabla \psi_\downarrow|^2   - \psi_\uparrow^*(i\partial_z\psi_\uparrow + \partial_y\psi_\downarrow)+ \psi_\downarrow^*(i\partial_z\psi_\downarrow + \partial_y\psi_\uparrow) +\psi_\uparrow^*\Omega\psi_\downarrow+\psi_\downarrow^*\Omega\psi_\uparrow
               + \frac{g}{2} (|\psi_\uparrow|^4 + |\psi_\downarrow|^4)  + g|\psi_\uparrow|^2 |\psi_\downarrow|^2 +  V(\bm{r},t) ({|\psi_\uparrow |}^2+{|\psi_\downarrow |}^2) ] d\mathbf{r}$.
By substituting the soliton-form trial solution into the above expression and employing the adiabatic approximate solution for the spinor parameters, we obtain:
$L = N_b (\mathbf{k} \cdot \dot{\mathbf{r}}_c
 + \textbf{A} \cdot \dot{\mathbf{k}})-E_s(\mathbf{k})- E_p(\mathbf{r}_c),
$
where $\textbf{A}=  i \begin{pmatrix}
C_1^*(\mathbf{k}) & C_2^*(\mathbf{k})
\end{pmatrix}
\dfrac{d}{d \mathbf{k}}
\begin{pmatrix}
C_1(\mathbf{k}) \\
C_2(\mathbf{k})
\end{pmatrix}$. This Lagrangian contains both kinetic action terms and energy terms associated with the evolving soliton. The kinetic action consists of two terms: the first term represents the classical action of the orbital motion of the soliton center, while the second one accounts for contributions from SOC.
The second term of the kinetic action is typically small compared to other terms in the Lagrangian, as it is proportional to the external force, which is assumed to be sufficiently weak to validate the adiabatic approximation used in the preceding deductions. Neglecting this small term reduces the equation to the  QPT equations. However, despite its small magnitude, this term contains essential topological singularities that may influence the global structure of quantum states (such as geometric phases) as well as soliton dynamics, as will be discussed below.

The Lagrangian equations of the above system  yield:  $\dot{\mathbf{k}}  = -\nabla V(\mathbf{r}_c)$,   and $\dot{\mathbf{r}}_c = \frac{1}{N_b} \nabla_{\mathbf{k}} E_s(\mathbf{k}) - \dot{\mathbf{k}} \times \mathbf{B}_m$, where $\mathbf{B}_m =\nabla_{\mathbf{R}} \times \boldsymbol{\mathbf A} = -\mathbf{R}/(2R^3)$ and $R = \sqrt{k_z^2 + k_y^2 + \Omega^2}$ (see details in \cite{SM}).
Compared to the motion equations derived from QPT, an additional term \(\dot{\mathbf{k}} \times\mathbf{B}_m\) appears. The coupling parameter \(\Omega\) can be formally treated as the x-component in a 3D momentum space. This term then corresponds to the Lorentz force in momentum space experienced by a moving soliton ``particle" in a VMM field \(\mathbf{B}_m\). The VMM is located at the origin of momentum space and has a charge of \(1/2\) \cite{Dirac,Berry1984}. The above motion equations for the soliton center are analogous to the semiclassical description of electrons in a periodic Bloch energy band, where the Berry curvature of a Bloch eigenstate acts as a VMM \cite{Niu}. In solid-state physics, this term is known to induce various anomalous electron transport phenomena \cite{Niu, Wang2021, Chaudhary2018, Zhu2024, Higuchi2024} and is responsible for the distinct topological properties of certain materials \cite{Niu6, Dalibard, Silberstein, Gianfrate2020}. This issue has also been discussed in terms of phase-space curvature in SOC ultracold atomic systems \cite{Armaitis2017}.

\begin{figure}
    \centering
    \includegraphics[width=80mm,height=65mm]{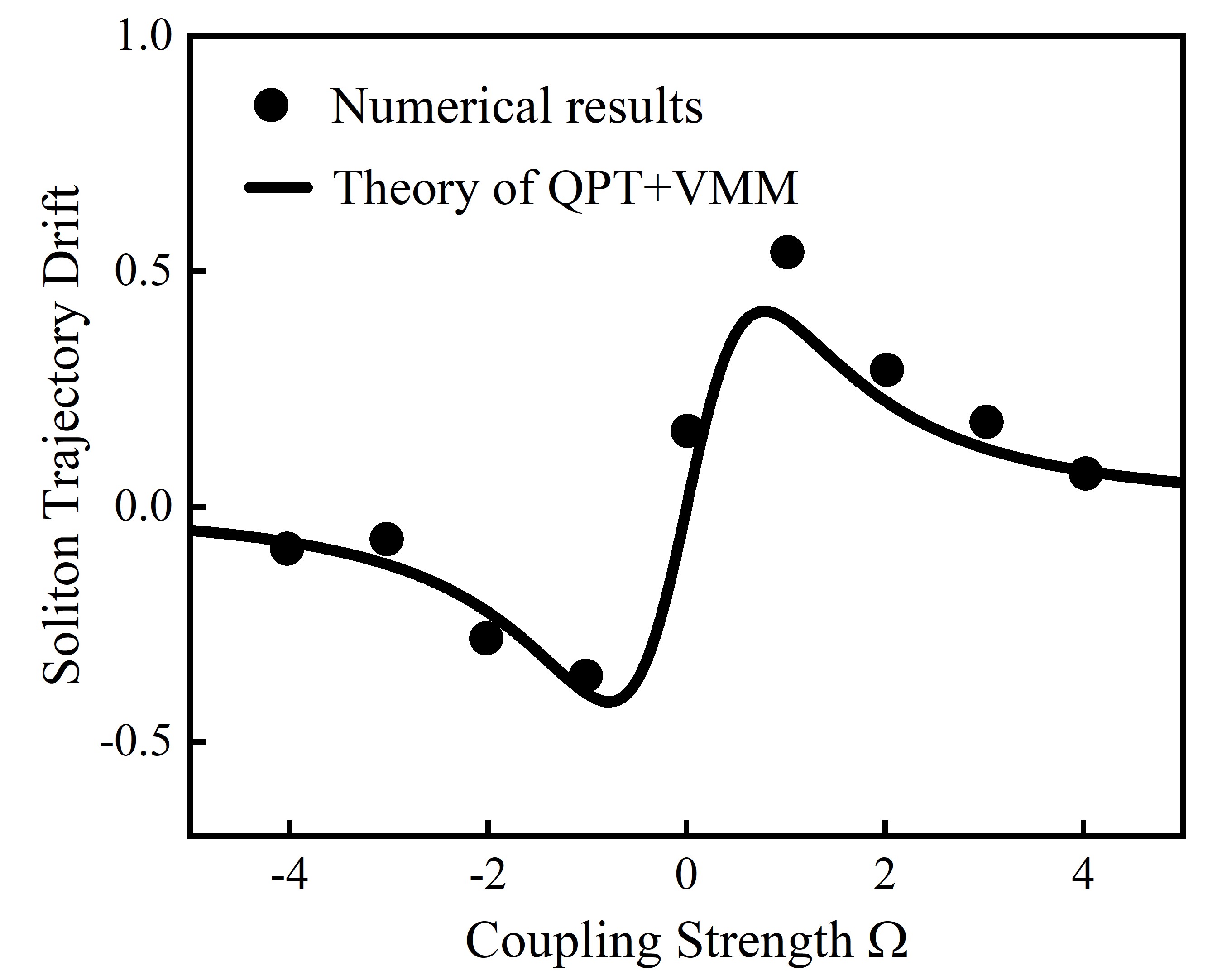}
     \caption{The maximum anomalous soliton trajectory drift during one period vs coupling strength $\Omega$.  }
    \label{fig2}
\end{figure}

We have numerically solved the above equations (i.e., QPT+VMM) and compared the results with those from QPT and GP simulations, as illustrated in Fig. \ref{fig1} (b) and (c). Our QPT+VMM results show good agreement with the numerical simulations of the GP equations, indicating that the VMM can completely compensate for the anomalous drift of the soliton orbit. The soliton trajectory drift depends on the coupling parameter $\Omega$. We have calculated the maximum drift values during one period and compared with the theory of QPT+VMM. As shown in Fig. \ref{fig2}, our theory agree well with the numerical simulation results. When $\Omega=0$, the drift tends to vanish since the $\dot{\mathbf{k}} \times \mathbf{B}_m$ always have zero components along the $y$ and $z$ directions in this case.

\emph{Geometric phase of the cyclic soliton.}
In addition to the anomalous orbital drift of solitons, the VMM possesses an intrinsic topological singularity that may influence the global structure of quantum states such as the geometric phase of cyclic motion \cite{Berry1984}. Cyclic evolution in physical systems is of interest both experimentally and theoretically \cite{AA, EschrigSpringer2011}, as it is associated with a phase that depends solely on the evolving wavefunction and is independent of the underlying Hamiltonian; this is termed the geometric phase \cite{Berry1984, AA}.
In the present work, we have observed perfect cyclic motion of solitons in a  SOC-BEC  system. In the following section, we will investigate the geometric phase associated with this cyclic evolution.

The time evolution equation for the total phase \(\theta\) of the soliton can also be derived using the Lagrangian variation method described above (see details in \cite{SM}): $\dot{\theta} =  \mathbf{k} \cdot \dot{\mathbf{r}}_c + \boldsymbol{\mathbf A} \cdot \dot{\mathbf{k}} - \mu$, where the chemical potential is  $\mu = \frac{E_s(\mathbf{k})  + E_p(\mathbf{r}_c) }{ N_b} + \frac{ga^2}{2 \ln 2}\left(\frac{2\ln 2}{3}-\frac{1}{6}\right)$.
In addition to the dynamical phase \(\theta_{\text{dyn.}}\), which corresponds to the time integral of the chemical potential, there exists a geometric phase expressed as:
$\theta_g = \oint \mathbf{k} \cdot d\mathbf{r}_c + \oint \boldsymbol{\mathbf A} \cdot d\mathbf{k}$.
The first term, arising from the action of the soliton's cyclic orbital evolution, corresponds to the non-adiabatic Aharonov-Anandan phase \(\theta_{\mathrm{A-A}}\) \cite{AA,Liu1998}. The second term represents the Berry phase \(\theta_{\mathrm{Berry}}\) \cite{Berry1984,Liu,Liu2010}, which originates from the adiabatic spin reversal driven by the weak segmented rotational force. According to Stokes's theorem, this term can be rewritten as \(\iint \mathbf{B}_m \cdot d\mathbf{S}\), i.e., the surface integral of the VMM field.

The phases associated with the cyclic soliton motion can be calculated by directly solving the GP equations numerically. The results of this calculation have been compared with the theoretical expressions derived above, as shown in Table I. Both the numerically observed phases and the theoretically calculated values are presented for three different coupling strengths of \(\Omega\). It is evident that the theoretical results are in good agreement with the observed values. Specifically, the Berry phases are well described by the flux of \(\boldsymbol{\mathbf B}_m\) enclosed by the closed curves in momentum space. Additionally, the motion equation for the soliton center clearly indicates that the Aharonov-Anandan phase does not generate any virtual fields in the parameter space, and its value can be modified by the trajectory drift associated with Berry phases.

\begin{table}
\caption{The observed total phases of soliton during one cyclic evolution and the corresponding theoretical values.}
\label{table:phase_measurements}
\begin{ruledtabular}
\begin{tabular}{cccccc}
\addlinespace[0.5em]
\multicolumn{1}{c}{} &
\multicolumn{1}{c}{Observed values} &
\multicolumn{4}{c}{Theoretical values} \\
\cline{3-6}
\cline{2-2}
\addlinespace[0.5em]
\multicolumn{1}{c}{$\Omega$} &
\multicolumn{1}{c}{$\theta_{\mathrm{total}}$} &
\multicolumn{1}{c}{$\theta_{\mathrm{total}}$} &
\multicolumn{1}{c}{$\theta_{\text{Berry}}$} &
\multicolumn{1}{c}{$\theta_{\text{A-A}}$} &
\multicolumn{1}{c}{$\theta_{\text{dyn.}}$} \\
\hline
\addlinespace[0.5em]
 1 & $10.70\pi$ & $10.60\pi$ & $-0.33\pi$ & $6.30\pi$ & $4.63\pi$ \\
 \addlinespace[0.5em]
 0 & $ 5.27\pi$ & $ 5.14\pi$ & $-\pi$      & $2.36\pi$ & $3.78\pi$ \\
 \addlinespace[0.5em]
-1 & $11.39\pi$ & $11.28\pi$ & $ 0.33\pi$ & $5.58\pi$ & $5.37\pi$ \\
\end{tabular}
\end{ruledtabular}
\end{table}

\emph{Conclusion.}
Taking two-component BEC as an example, we investigate the soliton dynamics modulated by the interplay between SOC effects and external field driving, and identify a notable VMM  that influences both the moving trajectory of a BEC soliton and the geometric phase acquired during its cyclic motion. Our theory can be readily extended to the multi-component cases \cite{39SP}, such as $^{87}$Rb with $F=1$, $^{23}$Na with $F=2$ and $^{52}$Cr with $F=3$,  in which the VMM with higher charges are expected.
These findings not only deepen our understanding of the role of quantum geometry in soliton dynamics, but also present a robust, experimentally accessible pathway to observe anomalous velocity drift, Berry phase, and Aharonov-Anandan phase effects through combining the soliton experiments \cite{Townes20,Townes21} and SOC-BECs experimental techniques \cite{Li2015,Lin2011}. On the other aspects, the strong nonlinearities are expected to induce distinct soliton dynamics, nonlinear geometric phases \cite{Liu,Liu2010} and other exotic phenomena that require further investigations.

\section*{Acknowledgments}
We are grateful to Tao Jiang and Haiyang Yu for their helpful discussions. L-C Zhao is supported by the National Natural Science Foundation of China (Contracts No. 12375005, No. 12235007 and No. 12247103). J Liu is supported by NSAF (Grant No. U2330401).


\end{document}